\documentclass[conference]{IEEEtran}
\IEEEoverridecommandlockouts
\usepackage{cite}
\usepackage{lipsum}
\usepackage{amsmath,amssymb,amsfonts}
\usepackage{algorithmic}
\usepackage{graphicx}
\usepackage{textcomp}
\usepackage{xcolor}
\usepackage{ifthen}
\usepackage{breakcites}
\usepackage{multicol, blindtext}
\usepackage{booktabs}
\usepackage{paralist,array}
\usepackage{url}
\usepackage{enumitem}
\usepackage{multirow}
\usepackage{float}
\def\BibTeX{{\rm B\kern-.05em{\sc i\kern-.025em b}\kern-.08em
    T\kern-.1667em\lower.7ex\hbox{E}\kern-.125emX}}
\usepackage{pdflscape}
\usepackage{afterpage}
\usepackage{caption}
\usepackage{array}
\usepackage{rotating}
\usepackage[utf8]{inputenc}
\usepackage[T1]{fontenc}
\usepackage{siunitx,threeparttable}
\usepackage[normalem]{ulem}
\newcommand{\PreserveBackslash}[1]{\let\temp=\\#1\let\\=\temp}
\newcolumntype{C}[1]{>{\PreserveBackslash\centering}p{#1}}
\newcolumntype{R}[1]{>{\PreserveBackslash\raggedleft}p{#1}}
\newcolumntype{L}[1]{>{\PreserveBackslash\raggedright}p{#1}}

\begin{document}

\title{On the Requirements for Serious Games geared towards Software Developers in the Industry}

\author{ \IEEEauthorblockN{Tiago Gasiba}
         \IEEEauthorblockA{\textit{Siemens AG} \\
         Munich \\
         \textit{tiago.gasiba@siemens.com}
} \and
         \IEEEauthorblockN{Kristian Beckers}
         \IEEEauthorblockA{\textit{Siemens AG} \\
         Munich \\
         \textit{kristian.beckers@siemens.com}
} \and
         \IEEEauthorblockN{Santiago Suppan}
         \IEEEauthorblockA{\textit{Siemens AG} \\
         Munich \\
         \textit{santiago.suppan@siemens.com}
} \and
         \IEEEauthorblockN{Filip Rezabek}
         \IEEEauthorblockA{\textit{Siemens AG} \\
         Munich \\
         \textit{filip.rezabek@siemens.com}}
}

\maketitle

\begin{abstract}
Teaching industry staff on cybersecurity issues is a fundamental activity that must be undertaken in order to guarantee the delivery of successful and robust products to market. Much research attention has been devoted to this topic over the last years. However, the research which has been done has not focused on developing secure code in industrial environments. In this paper we take a look at the constraints and requirements for delivering a training, by means of cybersecurity challenges, that covers secure coding topics from an industry perspective. Using requirements engineering, we aim at understanding the design requirements for such challenges. Along the way, we give details on our experience of delivering cybersecurity challenges in an industrial setting and show the outcome and lessons learned.
The proposed requirements for cybersecurity challenges geared towards software developers in an industrial environment are based on systematic literature review, interviews with security experts from the industry and semi-structured evaluation of participant feedback.
\end{abstract}


\section{Introduction}
In order to successfully deliver products to market, development in an industrial setting must follow existing laws, regulations and standards. Due to the increasing amount of successful hacking attacks, standardization bodies have been paying special attention towards introducing secure coding processes in the secure development lifecycle in the enterprise. Examples of standards that specifically mandate that companies, and therefore their software developer workforce, to follow secure coding guidelines and policies can be found, e.g. in~\cite{2015_PCI_DSS, 2012_BDEW, 2010_62443_2_1, 2018_62443_4_1, 2013_27002, 2010_NIST_800_37, 2013_BSI_Leitfaden_Webanwendungen, 2018_SAFEcode}.

Additionally to this, there is currently a strong driving force in the industry, researchers and even governments called {\it digitalization}. Towards this goal, in Vision~2020~\cite{siemens01_vision2020}, different industrial partners have come together and committed to a charter of trust~\cite{siemens02_charter}. This document outlines how the industry is willing to address the issues inherent with cybersecurity as a result of digitalization. One of the stated key principles is focusing on the dedication and effort that shall be spent towards professional cybersecurity education and training.

From many possible forms of training industrial staff in cybersecurity awareness~\cite{2014_Benenson_Defining_Security_Awareness}, we are particularly interested on training using Capture-the-Flag (CTF) exercises geared towards software developers.
Our motivation to use CTF as a form of awareness training comes from the work of Graziotin et al.~\cite{2018_Graziotin_Happy_Developers,2017_Graziotin_Unhapiness_SW}. Their work has investigated the fact that happy developers become better coders of software. CTFs are known to improve the happiness and satisfaction of the participants~\cite{2014_Davis_Fun_CTF,2019_Chothia}. Therefore we see an opportunity to, not only increase secure coding awareness, but also have a positive impact in the code produced by software developers.
However, we are only interested in CTFs that are openly available (e.g. open source) and are not part of a commercial solution. The main reasons we choose to take this approach is because openly available challenges and CTFs: allow to easily develop and adapt own challenges, allow free exchange of challenges with external partners, existing challenges have been more scrutinized due to their free availability and they lead to lowering the overall cost for delivering IT security awareness training through CTF.

Recently lot of work has been directed towards investigating how these serious games can be designed, built and deployed in order to deliver cybersecurity awareness trainings \cite{2015_Alexis_Renewed_Approach,2014_Mirkovic_Class_CTF,Hulin2017,2016_Namin_Teaching_Cybersecurity}. However, most of the work has not focused on the industry and its requirements.
Even more surprising, however, is the fact that no previous systematic requirement elicitation (based e.g. on Ghanbari et. al.~\cite{Hadi2015}) was found that addresses IT security training awareness through CTF-like serious games challenges for the industry.
This poses a great challenge since many serious games are being evaluated, but none are developed following a requirements engineering methodology. 
Our requirement elicitation methodology includes requirements from CTF-like events, which additionally gathers requirements from the CTF participants themselves.
Our requirements are from industry experts and therefore provide an excellent base line for other practitioners. Furthermore, we enable other practitioners with our requirements engineering methodology to elicit requirements for their individual CTFs. 

Davis et. al~\cite{2014_Davis_Fun_CTF} defines Capture-the-Flag events as having one of the following three types: {\it Attack-Defend}, {\it Attack-Only} and {\it Defend Only}. Typical topics covered in these CTFs range from web application security, cryptography, forensics, steganography, reverse engineering, mobile security and many other topics. Two prominent examples of Attack-Defend CTF are well-known (commercial) DEFCON CTF~\cite{DEFCON} and Hack-the-Box~\cite{HACK_THE_BOX}. Here the participants own some infrastructure and have two different tasks: that they need to protect it against adversaries and they need to attack the infrastructure of their opponents.
An example of Attack-Only CTF is the Jeopardy-style which involves the participants solving several questions and obtaining points for the correct solutions. In Defend-Only CTFs the participants are only given defensive challenges, typically having to secure infrastructure against attacks and maintaining its operations and functionality.

Participants of such events range from university graduates to professional penetration testers~\cite{2017_Alotaibi_Review_Gaming}. Our understanding is that these participants have generally a strong security background and sets of skills and are likely able to gather large practical experience from participating in several different CTF events~\cite{Svabensky2018,2014_Mirkovic_Class_CTF,Hulin2017}.

Preliminary results by Votipka~et~al. seem to suggest that CTFs can have a beneficial impact on secure software development~\cite{2019_Votipka_Hacking_CTF_Impact} through improved security thinking.
Votipka~et~al. work focuses on openly available CTFs which have an attacker perspective (jeopardy style).

In our work we try to understand the differences between existing CTFs and CTFs that can be used by software developers in the industry.
Given the three different types of CTF as defined by Davis~et.~al \cite{2014_Davis_Fun_CTF}, our initial assumption was that a defensive CTF would be the most adequate type in an industrial setting.
We would like to explore the requirements that the CTF challenges themselves must comply to in order to make them useful for software developers in the industry, e.g. by raising awareness on secure coding topics as mandated by standards.

Towards this goal, we have (1) conducted systematic literature review~\cite{2007_Kitchenham_SLR} of existing CTFs, (2) have asked IT security experts about their opinion on challenge design requirements and (3) have run four internal CTFs and gathered feedback using semi-structured interviews~\cite{1995_Drever_Semistructured_Interviews}.
The goal of running the CTF was originally intended as a preliminary validation of the challenge design requirements. We were able, however, to obtain additional design requirements from the gathered participants' feedback.

The CTF that we have run internally was based on existing and freely available open-source challenges and platform~\cite{2018_MBE_Cource,2018_CTFd_IO,2018_OWASP_JUICESHOP,2019_PCAP_REPO}.
The selected exercise categories have been based on OWASP~Top~10~\cite{2017_OWASP_TOP_10} and OWASP~IoT~Top~10~\cite{2017_OWASP_IoT_TOP_10}.

Section~\ref{sec:related_work} gives an overview on previously published and related work. Section~\ref{sec:methodolody} discusses the followed methodology. In section~\ref{sec:experiments} we briefly describe the three approaches and present some results. Based on the outcome of from previous steps, we summarize challenge design requirements in section~\ref{sec:derived_requirements}. In Section~\ref{sec:discussions} a critical analysis based on threat to validity is discussed. Finally a summary of the work and outline of further work is given in section~\ref{sec:conclusions}.
\section{Related Work}
\label{sec:related_work}

In this work we have specially paid attention to existing standards, literature on research methodology, on serious games and also on existing open source CTF challenges and supporting platform.

\subsection{Standards}
Software developers in the industry must follow the company internal policies and guidelines which are generally derived from existing standards and laws in alignment with business strategy, certification and accreditation.
IT security standards, such as \cite{2015_PCI_DSS,2012_BDEW,2010_62443_2_1,2018_62443_4_1,2013_27002,2010_NIST_800_37,2013_BSI_Leitfaden_Webanwendungen,2018_SAFEcode} mandate the implementation of secure coding guidelines. Companies should follow these standards on IT security in order to demonstrate due-diligence. However, although the standards mandate the implementation of secure coding, the specific guidelines are not defined in those standards. For secure programming in C, C++, Java, Perl and Android, there is a very good resource from Carnegie Mellon University~\cite{CERT_SEI_Coding_Standards} on secure coding guidelines.
For C\#~.Net, a good resource from OWASP can be found in~\cite{2019_OWASP_dotNet_CheatSheet}.

\subsection{Serious Games}
A lot of work has been recently devoted to the topic of Serious Games in education, in particular the Capture-the-Flag style of game. Some of the publications address the events themselves or the technique behind the challenges, e.g. in Švábenský~et~al~\cite{Svabensky2018} a competition is described whereby students taking a computer science course at the Faculty of Informatics in Brno create the challenges themselves (KYPO Cyber Ranges). Hulin~et~al~\cite{Hulin2017} propose a methodology of automatically creating new challenges by means of mutating existing ones and automatically performing bug injection. While this work can target software developers, it is also shown that not all the injected vulnerabilities are exploitable, leading to possible issues for the participants.
In~\cite{2014_Mirkovic_Class_CTF}, Mirkovic and Peterson investigate an adapted CTF method which they propose to enhance cybersecurity education for students. While their work includes on both attack and defend exercises, they focus and try to foster adversarial thinking. After the CTF takes place, the tutor explains possible solutions of the exercises. Additionally an in-class post-mortem analysis of the event helps students to identify their mistakes and therefore further improve their skills.

While many of these publications mostly take for granted the suitability of CTFs as a tool to enhance cybersecurity awareness~\cite{2014_Benenson_Defining_Security_Awareness}, this has been put into question~\cite{2016_Hendrix_Are_Games_Suitable,2014_Davis_Fun_CTF,2017_Alotaibi_Review_Gaming,2012_Cheung_Effectiveness_Competitions}.
In~\cite{2014_Chung_Learning_Obstacles}, Chung and Cohen evaluated several possible obstacles to effective learning through CTF. The major conclusions that they have arrived to are that the challenges need to be adapted to the participants, the difficulty level should be adequate to the participants and that the challenges should undergo a well defined design process.

Miljanovic et al.~\cite{2018_Miljanovic_Review_SeriousGames} have also reviewed CTFs targeting the ACM 2013 Computer Science Curricula guidelines and arrived to similar conclusions, in particular that the design aspects of such games need to be addressed properly. Serious games design aspects are analyzed, in a general form, in~\cite{2015_Alexis_Renewed_Approach}.
In~\cite{2017_Pesantez_Serious_Game_SLR}, Pesantez~et~al. perform systematic literature review based on~\cite{2007_Kitchenham_SLR} in order to understand serious game design methodology, frameworks and models.
Lameras~et~al.~\cite{2017_Lameras_Serious_Game_Design} have summarized in their study how design features can be planned, developed and implemented. Their work focuses on the learning outcomes, teacher roles, pedagogic value and game attributes.
Two foundational works which cover the design aspects of serious games can be found in~\cite{2016_Doerner_Serious_Games} and \cite{2016_Doerner_Enternainment_Computing}.

It is our belief that for a CTF event to be successful in the industry, it both needs to address the target audience and be designed using the best known design methodologies.
As a result of our literature research, we have found out that most of the work has been focusing on academic or IT security experts (e.g. pen-testers, network administrators, etc). We have found less publications related to CTFs geared towards software developers in the industry. 

In~\cite{2013_Radermacher_Gaps_Industry}, Rademacher gives some hints into the differences between academia and industry. Although their work does not focus on secure coding, we believe that their conclusions also extend to this area.

Oliveira~et~al.~\cite{2018_Oliveira_API_Blindspots} have identified that an additional factor why developers might write insecure code is based on application programming interface (API) blind spots. In their work, the authors have defined API blind spots as a lack of knowledge by the developer on the correct usage of programming APIs. They show how the misuse of these APIs can lead to unintentional software vulnerabilities.

\subsection{Open Source CTF Challenges and CTF Platform}
In order to have a head-start on IT security challenges, we have looked at \cite{2018_MBE_Cource,2018_CTFd_IO,2018_OWASP_JUICESHOP,2019_PCAP_REPO}. The Rensselaer Polytechnic Institute has published online their learning curricula on binary exploitation. These exercises have been taken as a basis to build challenges on secure coding guidelines for C and C++. The OWASP Juice Shop has been used as a basis for challenges on secure coding for web applications. The malicious PCAP repository was used as a source of challenges to test network forensics. This latter category is not part of our main goal. However, we have decided to include challenges on this category in order to (1) obtain feedback on additional challenges that are not targeting secure coding and (2) reduce participant frustration by allowing different challenges to be part of the CTF.

\section{Methodology}
\label{sec:methodolody}
In this paper, we unify academia and industrial research.  Towards this goal, we have used different research methodologies for academic and practitioner aspects. Along the way, we were supported by four security experts from  industry.
\subsection{Academia}
For the academia part of our work, we have decided to perform a lightweight version of systematic literature review (SLR) as defined by Kitchenham~\cite{2007_Kitchenham_SLR}. In our instance, we have performed the following steps:
\begin{enumerate}
    \item {\bf Planning the review}: select relevant databases and define search keywords
    \item {\bf Execution}: perform the online search and gather the results
    \item {\bf Analysis}: analyse the results and codify taking into consideration our research goal
\end{enumerate}

\subsection{Industry}
For the industry part of our work, we have performed semi-structured interviews based on open questions and feedback~\cite{1995_Drever_Semistructured_Interviews,2009_Harell_Semistructured_Interviews}. In our study we have decided to use a three-point Likert scale according to Jacoby~\cite{1971_Jacoby_3Point_Likert}.

At the beginning of every capture-the-flag event, a white-board discussion with all the participants was conducted. The goal of the discussions was to capture the expectations of the participants for the workshop, to describe how the game would be played and also to help with any setup issues. At the end of the event, the participants were given feedback forms where they could enter free text on what went good and what went wrong. Additionally, participants were given short evaluation questions based on a three-point Likert scale.
\section{Research and Experiments}
\label{sec:experiments}
To determine possible challenge design requirements (CDR) applicable for Capture-the-Flag (CTF) for industrial software developers, we have conducted our research based on the following principles: (1) systematic literature review (SLR) on possible CDR, (2) interviews with security experts and (3) semi-structured evaluation of feedback from CTF participants.
Not only did we base our work on traditional requirements engineering such as \cite{Pohl:2010}, but also on feedback from participants, following practice-oriented requirements elicitation approach.
In the following we present our results based on these three methods.

\subsection{Literature review}
As described in the methodology section, we selected for the literature review the following databases: Google Scholar, IEEE eXplorer, Springer and ACM Digital Library (see Table~\ref{tab:papers_slr}). Based on the authors' experience with IT security, we have decided to use the following search keywords: "serious games", "industrial ctf", "capture the flag", "design requirements" and "secure coding". In order to be considered relevant, screened publications for further consideration have been published between 2012 until 2019. Another criteria was that the papers to be considered should give details on learning aspects of CTF. This was necessary in order to make sure that CDR can be derived. Additional papers for consideration included those that addressed gaps between industry and academia and on the general topic of serious games (with focus on industry). Thesis, books, commercial flyers and posters were not considered. Also discarded were papers that based their work on simulation results, commercial CTFs and those that did not address challenge design aspects.
After reviewing, sorting and determining if the paper is to be included in our research, 11 papers have been selected. Table~\ref{tab:papers_slr} shows a list of the selected papers together with a short summary of their contents.
One surprising outcome of this step was the fact that we were able to find only a very limited number of papers that addressed capture the flag events for the industry.

\begin{table*}[ht]
  \centering
  \caption{Selected papers from systematic literature review}
  \label{tab:papers_slr}
  \begin{tabular}{|p{4.8cm}|L{1.5cm}|p{0.6cm}|L{9.5cm}|}
    \hline
      {\bf Paper} & {\bf Publisher}  & {\bf Year} & {\bf Short description}\\
    \hline
      \cite{2014_Chung_Learning_Obstacles}~Chung et al., Learning Obstacles in the Capture The Flag Model
      &
      USENIX
      &
      2014
      &
      This paper addresses problems in typical Capture-the-Flag events that lead to lowering of learning effect and skill improvement of participants. Among their conclusions are the fact that many CTF aim at testing {\it very obscure types of knowledge}. They also claim that difficulties in game playing, hint system, lacking quality assurance and infrastructure problems can also leads to a poor CTF experience.
      \\
    \hline
      \cite{2015_Alexis_Renewed_Approach}, Alexis et al., A renewed approach to serious games for cyber security
      &
      IEEE - International Conference on Cyber Conflict
      &
      2015
      &
      This paper argues that, while serious games have demonstrated pedagogic effectiveness, this has only happen in limited contexts. Further it claims that with proper design, the serious games could reach a much larger audience. In particular it details a methodology of serious games design for people with little to no knowledge on cybersecurity.
      \\
    \hline
      \cite{Svabensky2018} Švábenský et al., Enhancing Cybersecurity Skills by Creating Serious Games
      &
      ACM ITiCSE'18
      &
      2018
      &
      This work details the effectiveness of the KYPO Cyber Ranges conducted at the Institute of Computer Science at the Faculty of Imformatics in Brno, Czech Republic. This CTF is integrated as part of a computer science course, where students are given lectures on cybersecurity topics, have supervised practice and perform group work. This paper includes lessons learned from the CTF covering their successes and encountered problems.
      \\
    \hline
       \cite{2013_Radermacher_Gaps_Industry}~Rademacher et at., Gaps between industry expectations and the abilities of graduates
       &
       ACM SIGCSE'13
       &
       2013
       &
       This work, based on systematic literature review, explores the difference between what employees are expected to know in the industry and abilities that graduates obtain during their studies. The goal is to raise awareness on the gaps such that educators can better address them on their curriculum.
       \\
    \hline
      \cite{2016_Hendrix_Are_Games_Suitable} Hendrix et al., Game Based Cyber Security Training: are Serious Games suitable for cybersecurity training?
      &
      International Journal of Serious Games
      &
      2016
      &
      This paper investigates, using the systematic literature review methodology, the suitability of serious games for cybersecurity training. It claims that, although there are early indicators that this might be the case, this conclusion is not given. In particular, the paper addresses a gap between the target audience and the serious games challenges.
      \\
    \hline
      \cite{2018_Miljanovic_Review_SeriousGames} Miljanovic et al., A Review of Serious Games for Programming
      &
      Springer - Joint Intl. Conference on Serious Games
      &
      2018
      &
      In this work, the authors reviews literature on existing serious games for software programmers with a focus on the ACM 2013 computer science curricula guidelines. The paper also identifies a number of open problems in serious programming games. The main research questions are on which specific knowledge is covered and how are the games evaluated.
      \\
    \hline
      \cite{2019_Votipka_Hacking_CTF_Impact} Votipka et al., Toward a Field Study on the Impact of Hacking Competitions on Secure Development
      &
      The Workshop on Security Information Workers
      &
      2018
      &
      Initial results obtained by this publication indicate a positive effect on security thinking (i.e. culture), team communication and the handling of complex security problems as an impact of software developers participating in hacking competitions.
      The analyzed CTF have a offensive style and are shown, through feedback obtained by the researchers, to teach participants to think more as an attacker.
      \\
    \hline
      \cite{2018_Oliveira_API_Blindspots} Oliveira et al., API Blindspots: Why Experienced Developers Write Vulnerable Code
      &
      USENIX
      &
      2018
      &
      API blindspots are defined as misconceptions, misunderstandings or oversight by the developer, that can potentially lead to the introduction of security vulnerabilities into the developed software. The aim of this paper is partially to improve software development process by means of trainers addressing the identified findings.
      \\
    \hline
       \cite{2009_Colley_Sec_Code_Not_Enough}~J Colley, Why Secure Coding is not Enough: Professionals’ Perspective
       &
       Springer - Securing Electronic Business Processes 
       &
       2009
       &
       The paper outlines basic concepts that must be considered in a secure software development lifecycle. The paper takes a holistic approach and mostly addresses the fact that not all breaches are caused by vulnerable code. This work touches the points of software design problems (secure architecture). In this regard it explains that impact on overall security must be understood as also the interplay of different technologies
       \\
    \hline
      \cite{2017_Pesantez_Serious_Game_SLR} Pesantez et al., Approaches for Serious Game Design: A Systematic Literature Review
      &
      Computers in Education Journal
      &
      2017
      &
      In this paper, the authors have analyzed 51 studies on serious game design for the academia using systematic literature review. Several approaches to serious game design are identified. Furthermore this work summarizes general features of serious games and challenges. Additionally identified issues are described.
      \\
    \hline
       \cite{2017_Lameras_Serious_Game_Design} Lameras et al.,  Essential features of serious games design in higher education: Linking learning attributes to game mechanics
       &
       British Journal of Educational Technology
       &
       2016
       &
       This work aims at determining, using a systematic analysis, how serious games are conceptualized, modeled an researched. It also gives indications on possible learning attributes, game attributes, game categories, rules, roles, challenges and motivation.
       \\
    \hline
  \end{tabular}
\end{table*}

The next step in our research was to perform a coding step. Here, by carefully reading the selected publications, a total of 12 commonly challenge design requirements were inferred. This was done be tabulating the common patterns that have emerged in the publications that addressed some challenge design issue. The resulting entries in the table were then grouped into 12 CDRs and coded as challenge design requirements by three security experts.
Table~\ref{tab:requirements_from_slr} shows a summary of the results from this step, together with a list of the papers that support each requirement.

\begin{table}[ht]
  \centering
  \caption{Challenge design requirements from SLR}
  \label{tab:requirements_from_slr}
  \begin{tabular}{|p{5.5cm}|C{2cm}|}
    \hline
      {\bf Challenge Design Requirement} & {\bf Supported by}\\
    \hline
      1. Have a clearly defined learning goal  & \cite{2015_Alexis_Renewed_Approach,2013_Radermacher_Gaps_Industry,2018_Miljanovic_Review_SeriousGames,2014_Chung_Learning_Obstacles,2016_Hendrix_Are_Games_Suitable,2017_Lameras_Serious_Game_Design,2017_Pesantez_Serious_Game_SLR}\\
    \hline
      2. Adapted to background (job description) of participants & \cite{2015_Alexis_Renewed_Approach,2013_Radermacher_Gaps_Industry,2016_Hendrix_Are_Games_Suitable,2017_Lameras_Serious_Game_Design}\\
    \hline
      3. Well defined working mechanics (e.g. which tools to use or what to do) & \cite{2015_Alexis_Renewed_Approach,2018_Miljanovic_Review_SeriousGames,2017_Lameras_Serious_Game_Design}\\
    \hline
      4. Define and progressive level of difficulty & \cite{2015_Alexis_Renewed_Approach,2014_Chung_Learning_Obstacles,2017_Pesantez_Serious_Game_SLR}\\
    \hline
      5. Elicit discussions of the solutions (e.g. is there a better/simpler way to solve?) & \cite{2013_Radermacher_Gaps_Industry,2017_Lameras_Serious_Game_Design,2019_Votipka_Hacking_CTF_Impact,2018_Oliveira_API_Blindspots,Svabensky2018}\\
    \hline
      6. Provide possible solution after challenge solved & \cite{2014_Chung_Learning_Obstacles,2017_Pesantez_Serious_Game_SLR,Svabensky2018}\\
    \hline
      7. Adapted to the skill level of participants & \cite{2014_Chung_Learning_Obstacles,2017_Lameras_Serious_Game_Design,2017_Pesantez_Serious_Game_SLR}\\
    \hline
      8. Challenge includes hint that aid to arrive to the solution & \cite{2014_Chung_Learning_Obstacles,2017_Lameras_Serious_Game_Design}\\
    \hline
      9. Clear, standardized and simple solution (not based on obscure knowledge) & \cite{2014_Chung_Learning_Obstacles,2017_Pesantez_Serious_Game_SLR}\\
    \hline
      10. Planned duration of the exercise & \cite{2016_Hendrix_Are_Games_Suitable,2017_Pesantez_Serious_Game_SLR}\\
    \hline
      11. Explains issues arriving from interplay of different technologies or components & \cite{2009_Colley_Sec_Code_Not_Enough,2018_Oliveira_API_Blindspots}\\
    \hline
      12. Adapted to company internal secure coding guidelines and policies & \cite{2009_Colley_Sec_Code_Not_Enough}\\
    \hline
  \end{tabular}
\end{table}

After this step, the authors were surprised that so many CDR have been found.
This comes as even more surprising in light of the need of CDR opposed to the lack of work addressing it in academic publications.
This gave us a strong motivation to continue with our research. Since this input comes from academia background, we were specially interested in (1) evaluating the requirements together with practitioners and (2) knowing if the requirements can in fact be implemented in practice.

\subsection{Interview with security experts}
As part of our research, we have also conducted informal interviews with two security experts from the industry in order to understand their opinion and concerns on the design of secure coding challenges for software developers. The discussions with the security experts took place after the first CTF event took place.
The discussions were based on the following two questions:
\begin{itemize}
    \item According to your experience, how would you design a serious game challenge which is targeted for a software developer?
    \item How to design the challenges such as to better motivate software developers to follow established secure coding guidelines?
\end{itemize}

The conducted interviews and discussions with the security experts was based on three parts: 
\begin{itemize}
	\item in the first part, the experts gave their feedback to the posed questions
	\item after this, we have described the design requirements that we have obtained from systematic literature review and asked if the security expert agrees with it or not
	\item in the last step, the security experts have been asked if they would like to add more points to their answers from the first part
\end{itemize}

As a result of the second phase of the interview, the security experts have agreed on the challenge design requirements. Table~\ref{tab:expert_feedback} shows a summary of the feedback given by the security experts.

\begin{table*}[ht]
  \centering
  \caption{Expert feedback from interview and on challenge design requirements}
  \label{tab:expert_feedback}
  \begin{tabular}{|C{1cm}|C{12.5cm}|}
    \hline
      {\bf Security Expert} & \vspace{.3px}{\bf Feedback}\\
    \hline
    \#1
    &
    \vspace{-5px}
    \begin{compactitem}
    	\item Challenge aligned with company policies and secure coding guidelines and business objectives
        \item Clearly defined learning goal
    	\item Explain reason for security coding guidelines
    	\item Show consequences of vulnerabilities and its possible negative impact to the business
    	\vspace{-5px}
    \end{compactitem} \\ \hline
      \#2
      &
      \vspace{-5px}
      \begin{itemize}
      	\item Challenge should teach developer to avoid the most common mistakes
      	\item Challenge based on OWASP Top 10 and similar vulnerability databases
      	\item Categorize the challenges according to learning goals
      	\item Aligned with company policies and secure coding guidelines
      	\item Give solutions after the CTF event
      	\vspace{-5px}
      \end{itemize}
      \\
      \hline
  \end{tabular}
\end{table*}

\subsection{Capture-the-Flag platform and participant feedback}
In the last step of our research, we have implemented and deployed a CTF based on available open-source projects.
We have adapted some challenges towards the CDR and have also integrated other challenges that are not CDR compliant. The goal and the reason of this step was to gather feedback from the participants in order to understand the validity and practicability of the challenge design requirements in an industrial setting. Furthermore, this allows to validate the gathered theory from a practical point-of-view.

\subsubsection{CTF platform}
\begin{figure}[H]
	\centering
	\includegraphics[width=.95\columnwidth]{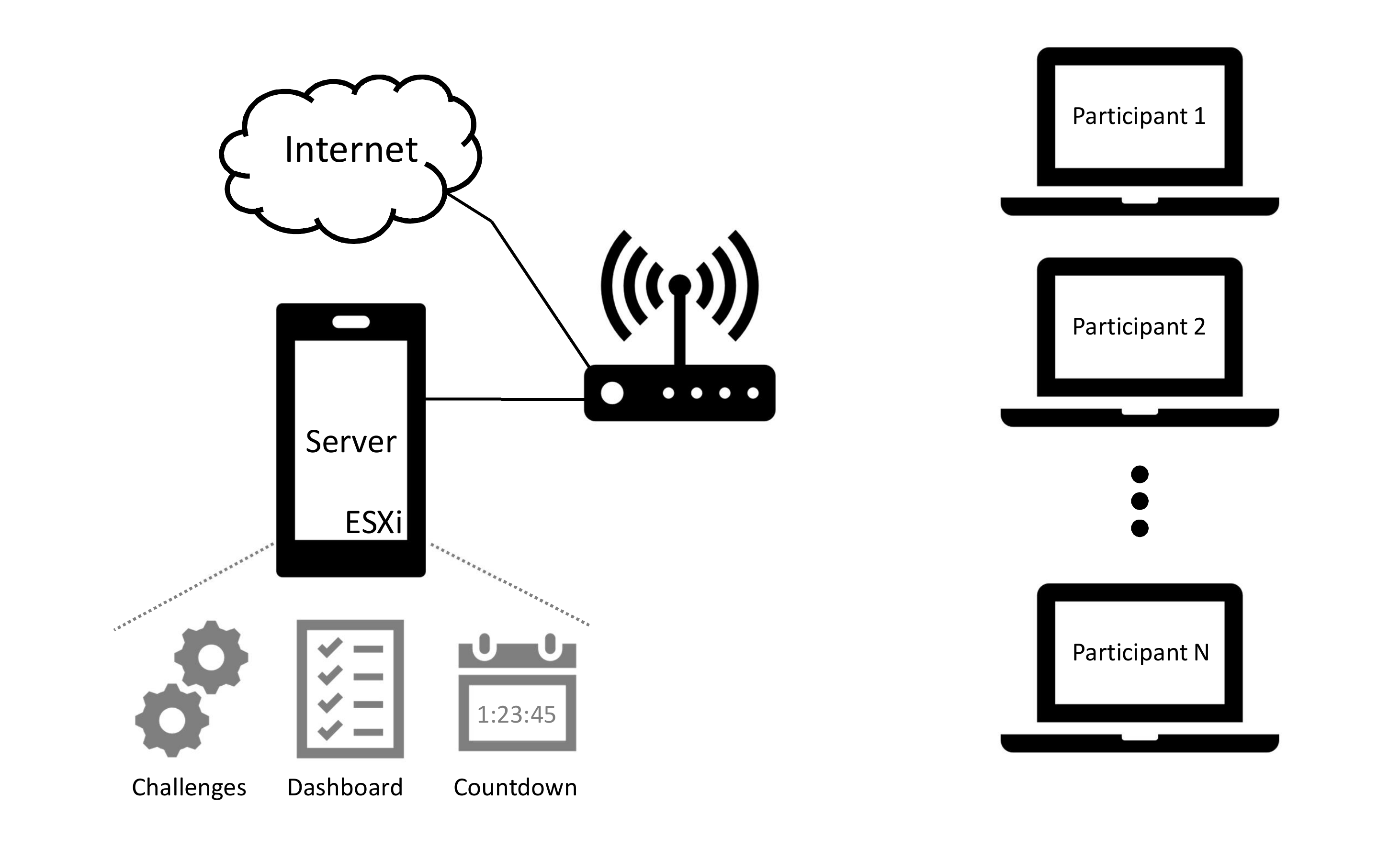}
	\caption{Capture-the-Flag Platform}
	\label{fig:ctf_deployment}
\end{figure}

Figure~\ref{fig:ctf_deployment} shows a simplified view of the deployment of the CTF platform.
Several different virtual machines and containers, which are set up on the main server, are used to host the CTF dashboard~\cite{2018_CTFd_IO} and also the different challenges:

\begin{itemize}
	\item {\it Web application} challenges based on~\cite{2017_OWASP_TOP_10}
	\item {\it Secure coding} challenges based on~\cite{2018_MBE_Cource}
	\item {\it Network forensic} based on~\cite{2019_PCAP_REPO}
	\item {\it Social engineering} based on self-developed questions
\end{itemize}

Additional hints were added, including how they work (e.g. which tools to use). They were classified according to difficulty level. Web application challenges were provided to web developers and the secure coding challenge were targeted at C/C++ developers. Network forensic and social engineering challenges were added, which are not part of the CDR.

The participants have both access to the internet, to enable searching for possible solutions, and also to the containers and dashboard.

\subsubsection{Evaluation of the CTF events}
We have performed four different CTF runs, as shown in table~\ref{tab:summary_CTF_events}. During all the runs, the authors have served in the role of security expert, giving advise on how to solve challenges during the game play, but also helping with eventual technical difficulties.

As briefly described in section~\ref{sec:methodolody}, a brief discussion and open questions were asked to the participants, before the events began. The purpose of the questions was to establish the expectations of the participants towards the event.
The following list summarizes the expectations expressed by the participants:
\begin{itemize}
    \item Train the ability to recognize security problems
    \item Exercises are related with daily work and practically oriented
    \item Enjoy the event and have fun
\end{itemize}

\begin{table}[H]
	\centering
	\caption{Summary of CTF Events}
	\label{tab:summary_CTF_events}
	\begin{tabular}{|c|c|c|c|c|c|}
		\hline
		{\bf Run} & {\bf Participants} & {\bf Type} & {\bf Category} & {\bf Nr.} & {\bf When}\\
		\hline
		1 & Security Experts    & Advanced & All & 11p & 2017 \\
		\hline
		2 & Software Developers & Basic    & Web & 12p & 2018 \\
		\hline
		3 & Software Developers & Advanced & Web & 6p  & 2018 \\
		\hline
		4 & Software Developers & Basic    & All & 30p & 2018 \\
		  & Pen Testers         &          &     &     &      \\
		\hline
	\end{tabular}
\end{table}

The first experiment, which took place in 2017 with 11 participants, was mostly geared towards assessing the developed CTF platform. Both security experts and junior students participated in this run. Feedback was gathered from open conversations after the event took place.

Since the feedback on platform stability was positive, this allowed us to perform a second and third experiment, this time with software developers of web applications from the industry. Both these events took place in the beginning of 2018 with 12 participants and 6 participants respectively.
The challenges for the first group were simple and had good hints, however for the second group the challenges were more difficult and the hints not so precise. The feedback obtained from the participants was gathered using open questions on what was good and what was bad about the experience.
The fourth run, which took place in middle 2018 with 30 participants, was performed with software developers (web and C/C++) and with pen testers. For this run, we have addressed the issues that have been reported in the previous runs. Feedback from the participants was gathered using (1) open questions and (2) three point Likert scale~\cite{1971_Jacoby_3Point_Likert} questions:
\begin{itemize}
    \item Did I learn something important?
    \item Would I recommend the CSC to other colleagues?
    \item The challenge difficulty adequate?
    \item Did I have fun?
    \item Did it fulfill my expectations?
    \item How was the duration of the event?
\end{itemize}

The following summarizes the main feedback we have obtained from the participants, related to CDR, during the second, third and fourth CTF run: {\it introduction to the exercises was very good, but many times it was not clear what to do in the exercise}; {\it the difficulty level was adequate and the support from the staff was welcome}; {\it the hints were helpful in solving some exercises, however some hints were either confusing or missing}; {\it the concept is very good, people had much fun during the CTF and the real-world examples are very good.}

However, we have also gathered some additional obtained feedback which was as expected, in particular that {\it knowledge of hacking tools should not be necessary in order to complete the challenges}.
One participant stated that: {\it [...] I found it great that the difficulty of the challenges increased [...] this way it wasn't so overwhelming at the beginning [...]}. This statement is very much in line with requirement \#4.
Other CDR, as shown in Table~\ref{tab:requirements_from_slr} have been partially validated by the participants.

The three-point Likert questions have also been analyzed.
Since the CTF was developed taking into consideration the challenge design requirements, as described previously, the goal of this analysis is to give indicators of validity that the participants still have an enjoyable experience~\cite{2018_Graziotin_Happy_Developers} while taking part of our CTF.

The participants that attended the CTF in the second, third and fourth run were told that the CTF event was a novel way of delivering awareness training in IT security for software developers.
As such, table~\ref{table_expecations_vs_recommend} shows the results of participants expectations vs. participant recommendation of CTF to other colleagues. Note that only about $53\%~(16)$ of the total number of participants that have attended the fourth CTF run have provided feedback.

\begin{table}[H]
  \centering
  \caption{Expectations vs Recommend}
  \label{table_expecations_vs_recommend}
  \begin{tabular}{|l|c|c|c|c|}
    \hline
      {\bf Fulfill Expectations} & {\bf NO} & {\bf NEUTRAL} & {\bf YES} & \\
      {\bf Recommend} & & & & \\
    \hline
                  & $(1)$ & $(0)$ & $(1)$     &                \\
      {\bf NO}  & $50\%$  & $0\%$ & $50\%$    & $100\%$        \\
                & $25\%$  & $0\%$ & $9,09\%$  &                \\
    \hline
                  & $(3)$ & $(1)$ & $(10)$    &                \\
      {\bf YES} & $21,43\%$ & $7,14\%$ & $71,43\%$ & $100\%$   \\
                & $75\%$    & $100\%$  & $90,91\%$ &           \\
    \hline
                & $100\%$ & $100\%$ & $100\%$ &                \\
    \hline
  \end{tabular}
\end{table}

We can clearly see that the majority of the participants had their expectations fulfilled and are willing to further recommend the awareness training format to other colleagues.
However, surprisingly, the software developers that are not willing to further recommend the CSC were not really interested if their expectations were fulfilled or not. We think that this effect is being observed due to the low amount of participant feedback. Also surprising is the fact that, participants which did not have their expectations fulfilled would still recommend the CTF to their colleagues. This table shows a clear tendency towards further recommendation, which gives us an indicator of overall happiness with the awareness training. 

\begin{figure}[h]
        \centering
        \includegraphics[width=.95\columnwidth]{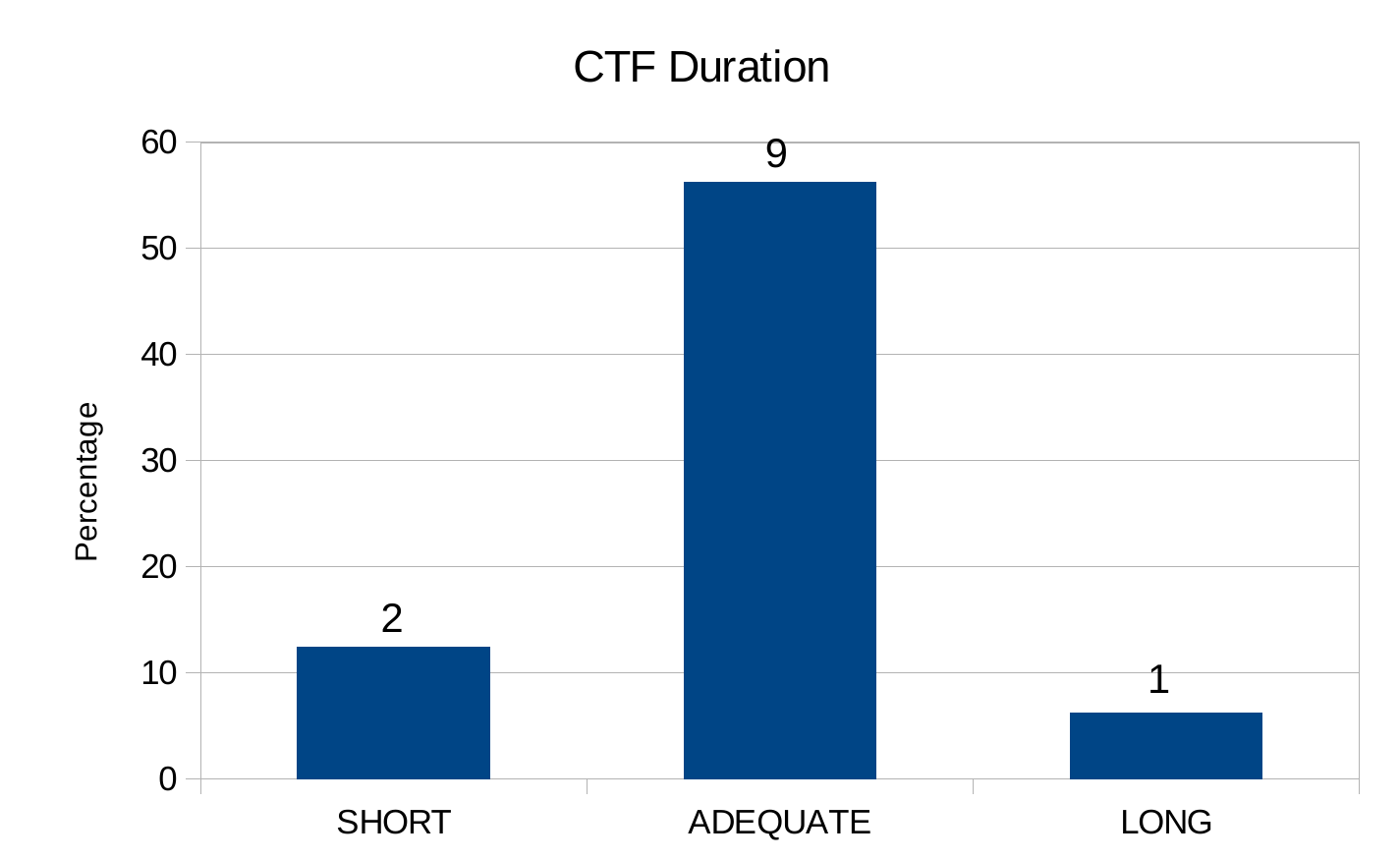}
        \caption{Results: CTF Duration}
        \label{fig:duration}
\end{figure}

Of particular sensitivity for the industry is the duration of the CTF event since the participants are not actively working for any company project. This means that the duration of training potentially penalizes a company twice: (1) by the amount of unproductive hours and (2) by the costs of the training itself.
We took as an assumption that the awareness training activities should take only one day. Therefore, with the exception of the first run, all the CTF events had the duration of one working day. Figure~\ref{fig:duration} shows feedback gathered from the participants on the suitability of this duration. We can observe that the majority of the participants has agreed that one day is an adequate duration.

\begin{figure}[h]
        \centering
        \includegraphics[width=.95\columnwidth]{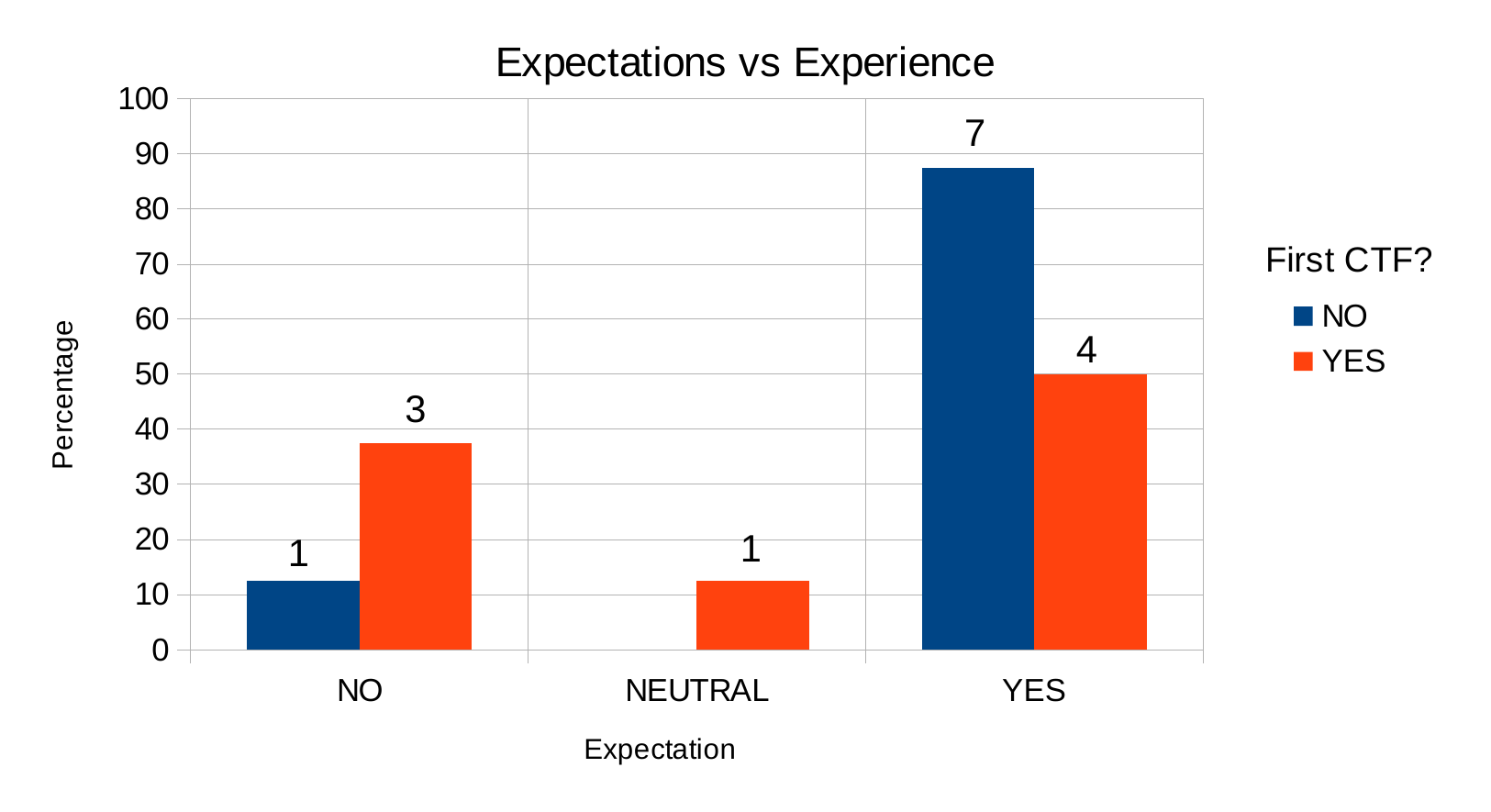}
        \caption{Results: Expectations vs Experience}
        \label{fig:expectation_vs_experience}
\end{figure}

Another important factor to consider is the fulfillment of the expectation of the software developers versus their experience on taking part in other similar kinds of competitions.
Figure~\ref{fig:expectation_vs_experience} shows that, for participants that had previous experience with CTF events, their expectations were moderately met; however, for participants that had no previous experience with CTF, they had mostly their expectations fulfilled. From our perspective, there are two surprising results: (1) we found a significant percentage of software developers that had already taken part in a similar competition and (2) there was a high amount of participants that had previously taken part in a similar CTF and did not see their expectations fulfilled.

We think that for the second point there is a bias introduced into the analysis due to the fact that in the fourth run pen testers also have taken part in the challenge. Due to the low-level barrier of the challenges, this group might be indicating that they have participated in other CTFs that were more interesting for their group.

Nevertheless, all the results hereby presented serve as a good indicator that the proposed format for CTF does fulfill participants expectations. We have also obtained from open discussions and feedback that the participants were happy during the event. These several factors serve as an indicator that the initial decisions taken for game design and that the CDR are validated.

Additionally, during the CTF events, we have collected some lessons learned and observations which are summarized as follows:

\begin{itemize}
	\item simple challenge hints are important for the players that wish to be more competitive and had already previous training in secure coding, however
    \item some participants require more than simple hints, e.g. complete description on the challenge, its background and impact on business; this helps lower the game play frustration and therefore increase the happiness of playing the game
    \item lessons learned from traditional CTFs is very low and their usability and applicability for software developers in an industrial context is also very low
    \item we could experience the excitement and the suitability of this kind of awareness training as a learning platform. Playing can be fun and an effective learning tool at the same time
    \item during in-game and off-game discussions, the software developers could find new ways and perspectives of viewing security issues, which helps to better understand the purpose of security coding guidelines
    \item the CTF can be used as a first learning tool on secure coding, but it is most likely better suited to reinforce already learned material
\end{itemize}

\section{Game Design Requirements}
\label{sec:derived_requirements}
Section~\ref{sec:experiments} has given extensive details on the three research phases that we have conducted in order to derive challenge design requirements. Table~\ref{tab:requirements_from_slr} shows a summary of the 12 CDR that have been inferred using SLR. In the same section, feedback from participants in one of the four CTF events have shown to corroborate with the established CDR. However, three additional CDR have resulted from both interviews with IT security experts and the feedback from participants of CTF events. These are the following:

\begin{enumerate}
	\setcounter{enumi}{12}
	\item developed challenges should focus on defender perspective
	\item challenges solutions should not require specific knowledge of hacking tools
	\item possible consequences of exploiting a vulnerability as a result poor secure coding guidelines practice should be mentioned in the challenge
\end{enumerate}

The complete list of CDR can be found in Table~\ref{tab:all_cdr_vs_ctf}.
The requirements presented in this table aid in the creation of well designed CTFs for the industry. Furthermore, as shown in the same table, it can be used to compare or benchmark existing CTFs towards the CDR and therefore assess their suitability for usage within the company.

\begin{table*}
\centering
\caption{Openly Available CTF Platforms vs Challenge Design Requirements for Industry}
\label{tab:all_cdr_vs_ctf}
\begin{tabular}{|m{3cm}|m{1.7cm}|m{1.6cm}|m{1.6cm}|m{1.9cm}|m{2.1cm}|m{1.7cm}|C{0.8cm}|}

      \hline
        {\bf Requirement} & {\bf AutoCTF~\cite{Hulin2017}}  & {\bf PicoCTF~\cite{2014_chapman_picoctf}} & {\bf PlaidCTF~\cite{2013_plaidctf}} & {\bf Class CTF~\cite{2014_Mirkovic_Class_CTF}} & {\bf CSAW CTF~\cite{2004_moses_teaching_security}} & {\bf KYPO Cyber Range~\cite{Svabensky2018}} & $\%$ \\
      \hline
        \vspace{2px}1. Have a clearly defined learning goal objective\vspace{2px}
        & 
        {\bf Not fulfilled} 
        & 
        {\bf Not fulfilled}
        & 
        {\bf Not fulfilled}
        & 
        {\bf Fulfilled}
        & 
        {\bf Fulfilled}
        & 
        {\bf Fulfilled}
        & 
        $50\%$
        \\
      \hline
        \vspace{2px}2. Adapted to background (job description) of participants developers\vspace{2px}
        & 
        {\bf Fulfilled}
        & 
        {\bf Not fulfilled}
        & 
        {\bf Not fulfilled}
        & 
        {\bf Not fulfilled}
        & 
        {\bf Not fulfilled}
        & 
        {\bf Not fulfilled}
        & 
        $17\%$
        \\
      \hline
        \vspace{2px}3. Well defined working mechanics (e.g. which tools to use or what to do)\vspace{2px}
        & 
        {\bf Not fulfilled}
        & 
        {\bf Not fulfilled} 
        & 
        {\bf Not fulfilled}
        & 
        {\bf Not fulfilled}
        & 
        {\bf Not fulfilled}
        & 
        {\bf Not fulfilled}
        & 
        $0\%$
        \\
      \hline
        \vspace{2px}4. Defined and progressive level of difficulty challenges\vspace{2px}
        & 
        {\bf Not fulfilled}
        & 
        {\bf Fulfilled}
        & 
        {\bf Fulfilled}
        & 
        {\bf Not fulfilled}
        & 
        {\bf Fulfilled}
        & 
        {\bf Not fulfilled}
        & 
        $50\%$
        \\
      \hline
        \vspace{2px}5. Elicit discussions of the solutions (e.g. is there a better/simpler way to solve?)\vspace{2px}
        & 
        {\bf Fulfilled}
        & 
        {\bf Fulfilled}
        & 
        {\bf Fulfilled}
        & 
        {\bf Fulfilled}
        & 
        {\bf Fulfilled}
        & 
        {\bf Fulfilled}
        & 
        $100\%$
        \\
      \hline
        \vspace{2px}6. Provide possible solution after challenge solved\vspace{2px}
        & 
        {\bf Not fulfilled}
        & 
        {\bf Not fulfilled}
        & 
        {\bf Not fulfilled}
        & 
        {\bf Not fulfilled}
        & 
        {\bf Not fulfilled}
        & 
        {\bf Not fulfilled}
        & 
        $0\%$
        \\
      \hline
        \vspace{2px}7. Adapted to the skill level of participants\vspace{2px}
        & 
        {\bf Not fulfilled}
        & 
        {\bf Fulfilled}
        & 
        {\bf Not fulfilled}
        & 
        {\bf Not fulfilled}
        & 
        {\bf Not fulfilled}
        & 
        {\bf Not fulfilled}
        & 
        $17\%$
        \\
      \hline
        \vspace{2px}8. Challenge includes hint that aid to arrive to the solution\vspace{2px}
        & 
        {\bf Fulfilled}
        & 
        {\bf Fulfilled}
        & 
        {\bf Fulfilled}
        & 
        {\bf Fulfilled}
        & 
        {\bf Fulfilled}
        & 
        {\bf Fulfilled}
        & 
        $100\%$
        \\
      \hline
        \vspace{2px}9. Clear, standardized and simple solution (not based on obscure knowledge)\vspace{2px}
        & 
        {\bf Not fulfilled}
        & 
        {\bf Fulfilled}
        & 
        {\bf Not fulfilled}
        & 
        {\bf Fulfilled}
        & 
        {\bf Not fulfilled}
        & 
        {\bf Fulfilled}
        & 
        $50\%$
        \\
      \hline
        \vspace{2px}10. Planned duration of the exercise\vspace{2px}
        & 
        {\bf Not fulfilled}
        & 
        {\bf Not fulfilled}
        & 
        {\bf Not fulfilled}
        & 
        {\bf Not fulfilled}
        & 
        {\bf Not fulfilled}
        & 
        {\bf Not fulfilled}
        & 
        $0\%$
        \\
      \hline
        \vspace{2px}11. Explains issues arriving from interplay of different technologies or components\vspace{2px}
        & 
        {\bf Not fulfilled}
        & 
        {\bf Fulfilled}
        & 
        {\bf Fulfilled}
        & 
        {\bf Fulfilled}
        & 
        {\bf Fulfilled}
        & 
        {\bf Fulfilled}
        & 
        $83\%$
        \\
      \hline
        \vspace{2px}12. Adapted to company internal secure coding guidelines and policies\vspace{2px}
        & 
        {\bf Not fulfilled}
        & 
        {\bf Not fulfilled}
        & 
        {\bf Not fulfilled}
        & 
        {\bf Not fulfilled}
        & 
        {\bf Not fulfilled}
        & 
        {\bf Not fulfilled}
        & 
        $0\%$
        \\
      \hline
        \vspace{2px}13. Challenges are put from the defensive perspective\vspace{2px}
        & 
        {\bf Not fulfilled}
        & 
        {\bf Not fulfilled}
        & 
        {\bf Not fulfilled}
        & 
        {\bf Not fulfilled}
        & 
        {\bf Not fulfilled}
        & 
        {\bf Not fulfilled}
        & 
        $0\%$
        \\
      \hline
        \vspace{2px}14. Solutions does not require specific knowledge of hacking tools\vspace{2px}
        & 
        {\bf Not fulfilled}
        & 
        {\bf Not fulfilled}
        & 
        {\bf Not fulfilled}
        & 
        {\bf Not fulfilled}
        & 
        {\bf Not fulfilled}
        & 
        {\bf Not fulfilled}
        & 
        $0\%$
        \\
      \hline
        \vspace{2px}15. Challenges should raise awareness on possible consequences of malicious attack\vspace{2px}
        & 
        {\bf Not fulfilled}
        & 
        {\bf Not fulfilled}
        & 
        {\bf Not fulfilled}
        & 
        {\bf Fulfilled}
        & 
        {\bf Not fulfilled}
        & 
        {\bf Not fulfilled}
        & 
        $16\%$
        \\
       \hline~
        ~~~~~~~~~~~~~\% 
        & 
        \vspace{2px}{\it ~~~~~$20\%$}\vspace{2px}
        & 
        \vspace{2px}{\it ~~~~~$40\%$}\vspace{2px}
        & 
        \vspace{2px}{\it ~~~~~$26\%$}\vspace{2px}
        & 
        \vspace{2px}{\it ~~~~~$40\%$}\vspace{2px}
        & 
        \vspace{2px}{\it ~~~~~$33\%$}\vspace{2px}
        & 
        \vspace{2px}{\it ~~~~~$33\%$}\vspace{2px}
        &        
        \\
      \hline
    \end{tabular}
\end{table*}

\subsection{Existing Capture-the-Flag events vs Requirements}
\label{sec:ctf_vs_requirements}
Based on the inferred 15 challenge design requirements, we would like to check if openly existing CTFs can be suited for awareness training on secure coding for software developers in the industry.
Towards this goal we have looked at~\cite{CTF_TIME} and cross referenced with the literature we have previously reviewed in this work. The large amount of existing CTFs together with the difficulty of obtaining specific information about them does not make it possible to analyze them all. We have therefore decided to choose 6 different CTFs which, not only we think are representative of the whole existing CTFs, but are also described in the literature: AutoCTF~\cite{Hulin2017}, PicoCTF~\cite{2014_chapman_picoctf}, PlaidCTF~\cite{2013_plaidctf},  Class CTF~\cite{2014_Mirkovic_Class_CTF}, CSAW CTF~\cite{2004_moses_teaching_security} and KYPO Cyber Range~\cite{Svabensky2018}.
According to our experience and available information about these six CTFs, we have mapped the applicability of every CDR to the CTF. We have categorized the matching between the CTF and the requirement as {\it Not Fulfilled} and {\it Fulfilled}. This decision was based on both the authors' experience and publicly available information. The decisions were also confirmed by three additional IT security experts from the industry. The results are summarized in Table~\ref{tab:all_cdr_vs_ctf}. On the last column of the table, the percentage of fulfilled requirements is shown for each requirement and on the last row, the percentage of fulfilled requirements for each CTF is also shown.

In this table we have also decided to look at the percentage of CDR fulfillment. This value serves just as an indicator of the number of fulfilled CDR for a given CTF: lower values mean inadequacy for the industry while higher values mean adequacy for the industry, according to our CDR. It can also be used to compare two different CTF based on fulfillment rate.
Our expectation, according to our experience, was that, although the CDR defined in this work are specific for the industry, a given minimum threshold of about $80\%$ fulfillment of CDR would be achieved by many of the investigated CTFs. However, we found out that this was not the case.

Table~\ref{tab:all_cdr_vs_ctf} shows that existing open source CTFs are not adequate for performing awareness training on secure coding for software developers in the industry, since many of the challenge design requirements are not fulfilled at all.
Although still with a low value, not surprisingly, PicoCTF and Class CTF show as the two best solutions with about $40\%$ CDR fulfillment each.

None of the analyzed CTFs fulfills Req~\#3, Req~\#6, Req~\#10, Req~\#12, Req~\#13 and Req~\#14.
The reason why these requirements are not fullfiled are the following:
\begin{itemize}
    \item {\it Missing Req~\#3}: on all the CTFs, the participants need to find out by themselves how the challenge is working and are not given a head-start
    \item {\it Missing Req~\#6}: we could not find any information that, after solving a certain challenge, feedback on the correct solution is given for any of the analyzed CTFs
    \item {\it Missing Req~\#10}: no claim was found for any of the six CTFs stating that the challenges were developed in such a way as to be solvable by the participants in a given amount of time
    \item {\it Missing Req~\#12}: none of the CTFs combines and integrates directly in their challenges a clear pointer to which secure coding guideline was not followed 
    \item {\it Missing Req~\#13}: on all the analyzed CTFs, none mentions that the challenges should be solved using defensive and secure software development strategies
    \item {\it Missing Req~\#14}: due to the existing attacker-perspective by the analyzed CTFs, specific hacking tools are used and even encouraged in order to solve the challenges
\end{itemize}

Note that on table~\ref{tab:all_cdr_vs_ctf}, some CDR  (namely 3, 6 and 10) are not fulfilled by any of the CTFs that the authors have selected for comparison. This should not be taken in any way as a bad quality indicator for such CTFs
This is rather the result of the author's own experience and publicly available information by the date of publication. Also, we think that the main reason for this disparity lies in the fact that the CTFs have not been developed according to the CDR and are not targeting the industry. The CTFs are very well suited, adequate and effective for the purpose they were developed for.

In particular, for Req~\#6, it is our experience (which was confirmed many times by participant feedback) that solutions to the exercises should be provided at the end of the event. This allows the learning effect of the game to be maximized, while lowering frustration. The main reasons for this are the following:
\begin{itemize}
    \item Allows participants to review the exercises after the event
    \item Provides notes that can be used as reminder
    \item Can show and give a different solution
\end{itemize}

\section{Impact of this work}
\label{sec:impact}
The main contributions of this work is summarized in Table~\ref{tab:all_cdr_vs_ctf}, were we can find 15 CDRs for CTF in the industry.
Additionally we give a comparison table of existing CTFs and show their possible weaknesses. Even if the target of the individual CTF is not the industry, it is hoped that this work can foster further investigation and improvement of the existing CTFs, or creation of new ones, according to the CDR.
Also, in case one of the CTFs compared in this work (or even a different one) are being used in an industrial environment, this table helps to have a critical look at the game and either give an impulse to change to a different one or to improve on the existing.
Another contribution of this work is on the used methodology for requirements elicitation, which corroborates with~\cite{Hadi2015}.
In particular, the requirement elicitation method can be re-used to derive requirements that are more aligned with individual companies.
\section{Discussions on validity}
\label{sec:discussions}
In this paper we have derived 15 challenge design requirements for CTFs geared towards raising secure coding awareness for software developers in the industry.
We have conducted systematic literature review, from academia perspective, resulting in the identification of 11 relevant papers. We also interviewed IT security experts in the industry. Four different runs of CTF with participants from the industry have been performed in order to validate the findings from a practitioners point-of-view.
The participants of the CTF events also played a crucial role in their contribution to requirements elicitation from a practitioner's point-of-view.

\subsection{Threat to validity}
In our work we can see the following possible threats to the validity of our conclusions:

\begin{itemize}
	\item our work might have left out some relevant research paper(s) that might lead us to different conclusions

	\item feedback gathered from participants was not based on previously developed questionnaires. Most of the feedback obtained from participants was done with open questions and discussions and therefore the participants might have forgotten to tell us some important facts

	\item the amount of participants that took part on the gathered feedback was relatively low and therefore the statistical relevancy of the data needs to be further investigated
	
	\item since the last CTF event was an in-house open event, two pen testers have participated and also provided feedback. their collected answers might introduce bias into the data
	
	\item since not all CTFs have detailed publicly available information, our matching of CDR to existing CTF might miss-categorize some points

\end{itemize}

We are however convinced that, due to our experience in IT security in the industry, that the main ideas presented in this work are according to our practical observations.

\section{Conclusion and Further Work}
\label{sec:conclusions}
Cybersecurity is gaining more attention over the last years, be it through increasing cybercrimes and their impact in the industry, be it through the development of the cybersecurity offerings from different companies, through the news, through legislation and standards, etc.
In order to deliver robust and secure products to market, the industry needs not only to address corporate security but it specially needs to train their software developers to write secure code.
Due diligence implies that one of the main concerns from the industry is on how to proper train its software developers in an efficient, effective, relevant and accurate manner on secure coding.
The traditional solution to address this issue has been to train software developers using a class-room methodology. However, new methodologies are emerging that promise higher information retention but also added participant satisfaction and increased compliance with secure coding guidelines and policies. One of these methodologies is using a Capture-the-Flag (CTF) style game. However, CTFs have been designed to train pen testers and white hat hackers and have not been designed and adapted to teach IT security to software engineers on secure coding topics. 
There is extensive literature on design aspects of the CTF, but there is little information on the challenge design requirements (CDR) for CTFs geared towards software developers, specially for the industry.
This means that scientific work has been using and reporting CTFs, but these have not been designed taking requirements engineering into consideration.

In this work, we took a look at how to arrive at CDRs for such kind of CTF. Our research was done partially with academic background and partially in the industry.
In particular, we have not only done literature reviews (based on Kitchenham's approach) and gathered feedback from security experts, but we have also used CTF participants themselves as input for our requirements elicitation.
We have chosen to take an approach with open source software, due to the fact that it allows cost reduction for delivering CTF events, the challenges have been scrutinized by the open source community, allows the free exchange of challenges with external partners while also allowing in-house development and adaptation of such existing challenges.

During four different CTF runs, feedback from the participants has been gathered using semi-structured approach and questions using a simple three-point Likert scale.
The purpose to gather participant feedback was to validate the challenge design requirements obtained from the systematic literature review. However, during this phase, we have discovered new requirements which were missing in the first phase of the research. Furthermore, we were able to confirm that defensive challenges as defined by Davis~et.~al are more appropriate for industrial environments.

While many of the results have been expected, there were some new unexpected insights. In particular, we found out that, according to our expected threshold for fulfilment rate of CDR, the analysed CTFs only achieved max $40\%$. We attribute this to the fact that the CTFs are, until now, not designed for the purpose we are seeking and also not with the requirements gathered in this work.
We have also shortly described possible threats to validity of our work.

The results hereby obtained encourage us to continue researching in this area and direction.
In particular, with this work we hope to improve the design of future CTFs, even if not with an industry target. We also hope to set some comparison measures between CTFs on their suitability for industrial environments.

As further work, we propose a critical evaluation of the CDR, in particular to understand if among the 15 CDR there are any conflicts.
Our next goal is to use action design research methodology to improve the CTF artifact to provide awareness training on secure coding for software developers in an industrial context. We also plan to collect CTF data on CTF player behaviour so as to identify different player types with the goal of aiding in understanding secure coding practices.
\section*{Acknowledgements}
The authors thank the participants of the CSC Workshops and the security experts at Siemens for the useful feedback and discussions.
The authors also thank Dr. Dreger from Siemens CT for all the encouragement, support and fruitful discussions and the anonymous reviewers for their constructive feedback that has significantly improved this work. 


\end{document}